\documentclass{article} 
 \usepackage{authblk}
\usepackage{graphicx}
\usepackage{hyperref}
\usepackage{authblk}
\usepackage{orcidlink}
\usepackage{abstract}
\emergencystretch=\maxdimen
\def\updfevolv{{\sc uPDFevolv2}}
\title{The role of the intrinsic-$k_{\rm T}$ and soft gluon contributions in Drell-Yan production 
\date {}
\thanks{Presented at  Hadron Structure and Fundamental Interactions 2024 (HSFI 2024)}} 
\author{Nata\v sa~Rai\v cevi\' c on behalf of the CASCADE Group}
\affil{University of Montenegro, Faculty of Science and Mathematics\\
 Podgorica, Montenegro}
\begin{document}
\maketitle
\begin{abstract}
A review of the most recent results on determination of parton internal transverse momentum in  initial colliding hadron obtained within the Parton Branching Method is presented and its interplay with non-perturbative soft gluon contribution in Drell-Yan pairs production is discussed.
\end{abstract}

\label{sec:intro}
\section{Introduction}

The Drell-Yan (DY) opposite sign lepton pairs get sizable transverse momenta through higher order processes manifested via QCD initial state radiation (ISR). The low transverse momentum partons give contribution to the low DY pair transverse momentum distribution and non-perturbative processes such as internal transverse motion of partons  and soft gluon emissions which need to be resummed are the main contributions in this kinematical region. Therefore, the study of the DY production cross section in hadron-hadron collisions  as a function of the pair's transverse momentum is crucial for understanding non-perturbative processes and a powerful tool for disentangling them. 

The first part of this paper is a review of the results on the determination of the internal transverse momentum of partons, the intrinsic-$k_{\rm T}$,  and its interplay with the  contribution of the soft gluon, especially its non-perturbative part, using Parton Branching (PB) Method. The intrinsic-$k_{\rm T}$ as a function of the collision energy will explain the interplay between the two processes.  These results are  followed by  an example from PYTHIA Monte Carlo event generator that confirms the findings obtained by the PB Method.      

\label{sec:pb}
\section{Non-perturbative contributions to the DY production studied by PB Method }

The PB Method~\cite{h1,h2} is based on the Transverse Momentum Dependent (TMD) parton distribution function, ${\-\cal A}_a(x,{\bf k_{\perp}}, \mu^2)$, which gives the probability that a parton $a$ inside the initial hadron has a transverse momentum {\bf k$_\perp$}, and a longitudinal momentum that is the $x$ fraction of the momentum of the initial hadron on a certain evolution scale, $\mu$, and in detail is described in~\cite{pb}. 
The partons with very small transverse momenta, whose treatment remains problematic in standard MC generators, are well controlled via TMDs. The soft contributions originate mainly from two 
non-perturbative processes: the internal transverse motion of the parton inside the colliding hadron through which the parton acquires an initial transverse momentum, commonly called intrinsic-$k_{\rm T}$, and soft gluon emissions which need to be resummed.    

The intrinsic-$k_{\rm T}$ is introduced through the TMD at starting scale $\mu_0$, ${\cal A}_a(x,{\bf k_{\perp,0}}, \mu_0^2)$,  which is parametrised by a collinear Parton Distribution Function (PDF) at the initial scale, $f_a(x,\mu_0)$, and a Gaussian distribution with width $\sigma$ and zero mean where the width is a measure of the internal transverse momenta and is related to the parameter $q_s$ used in the PB Method as $\sigma ^2 = q_s^2/2$:
%%%%%%%%%%%%%%%
\begin{equation}
 { {\cal A}}_a(x, k_{\perp 0},\mu^2_0) = f_a(x, \mu_0^2) \cdot {\rm exp}(-|k_{\perp 0}^2|/q_s^2)/\sqrt{(\pi q_s^2)}
 \end{equation}      

The soft gluon contribution is mainly contained in Sudakov form factor which gives the probability that there is no resolvable emission  between the two evolution scales. If we denote by $z$ the fraction of the parton's longitudinal momentum that is transferred at the branching and by $z_{\rm M}$ the parameter that defines the upper limit of $z$ below which the branching is still resolvable, the Sudakov form factor for the event evolving from scales $\mu_0$ to $\mu$ is given by the following expression:
\\
\begin{equation}
\label{sud-def}
 \Delta_a ( z_M, \mu^2 , \mu^2_0 ) = 
\exp \left(  -  \sum_b  
\int^{\mu^2}_{\mu^2_0} 
{{d {\bf q}^{\prime 2} } 
\over {\bf q}^{\prime 2} }
%\nonumber\\    
 \int_0^{z_M} dz \  z 
\ P_{ba}^{(R)}\left(\alpha_s , 
 z \right) 
\right).   
\end{equation}
\\
where $P_{ba}^{(R)}\left(\alpha_s , z \right)$ is  the splitting function for splitting parton $b$ into parton $a$ with emission of a parton $c$ with transverse momentum $q_{\rm \perp}$ in the branching.  
It is shown that $z_{\rm M} \rightarrow 1$ provides the exact solution of the DGLAP evolution and that the angular ordering provides the independence of the TMDs from the value of $z_{\rm M}$ when it is around 1~\cite{h2}. According to the angular ordering, the scale is evaluated as $q' = q_{\rm \perp} / (1 - z)$. 
Evolution of TMD sets is generated with \updfevolv \cite{updf}.
Analysis of hadron-hadron collisions within the PB Method are based on the TMD set called PB-NLO-2018 set2~\cite{pb} which is obtained by choosing $q_{\rm \perp}$ for the scale at which strong coupling is evaluated, which follows from
angular ordering, $\alpha_s(q^2_{\rm \perp}) = \alpha_s(q'^2(1-z)^2)$. Thus, there is a value of the minimal transverse momentum of the emitted parton at the branching, $q_0$, which separates the perturbative ($q_{\rm \perp} > q_0$) from the non-perturbative part ($q_{\rm \perp} < q_0$) where strong coupling is frozen at $q_0$. This introduces the intermediate $z$-scale, $z_{\rm{dyn}} = 1 - {\frac {q_0} {q'}}$ and two regions can be defined over the $z$-interval: $0 < z < z_{\rm {dyn}}$ for the perturbative region and  $z_{\rm {dyn}} < z < z_{\rm M}$ for the non-perturbative region. In the default settings of the PB Method, $z_{\rm M} \simeq 1$ to include as many soft emissions as possible. According to this, the integral in the exponent of Sudakov form factor can be split into two integrals in $z$, so that the Sudakov form factor~(\ref{sud-def}) can be expressed as the product of a perturbative form factor and a non-perturbative form factor~\cite{gluon}:

\begin{equation}
 \Delta_a ( z_M, \mu^2 , \mu^2_0 ) 
= \Delta_a^{(P)} ( z_M, \mu^2 , \mu^2_0 ) \cdot \Delta_a^{(NP)} ( z_M, \mu^2 , \mu^2_0 )
\end{equation}
 
A non-perturbative parameter, $q_s$, describing the intrinsic-$k_{\rm T}$ distribution has to be determined from experimental data. To compare the predictions with data, CASCADE3 Monte Carlo event generator based on the TMDs provided by the PB Method was used~\cite{cascade} in which the transverse momentum of the hard process was determined by the TMDs and the DY production at NLO was simulated by MADGRAPH5$\_$AMC@NLO~\cite{madgraph}. The final predicted cross section distributions to be compared with the experimental data were obtained using the Rivet tool~\cite{rivet}.  

The determination of the intrinsic-$k_{\rm T}$ width, $q_s$, was performed using the latest  available measurements  of the DY pair production cross section as a function of the pair transverse momentum over a wide DY invariant mass range obtained by the CMS Collaboration~\cite{CMS} which provided the detailed uncertainty breakdown contained in the covariance matrix used to calculate and minimise $\chi^2(q_s)$ obtained by comparing the data with CASCADE3 predictions.  The obtained value of the width and its uncertainty is $q_s = 1.04 \pm 0.08$~GeV~\cite{ktpaper}. 

For cross-checking the obtained result and getting the dependence of $q_s$ on the invariant mass of the DY pair and the collision energy, $\sqrt s$, other available experimental measurements were used and it was shown that there is no or very weak dependence of $q_s$ on the invariant mass and collision energy~\cite{ktpaper} . 

\label{sec:pes}
\section{Energy scaling of the intrinsic-$k_{\rm T}$ }

In contrast to the result obtained by CASCADE3 using TMDs, shower based Monte Carlo event generators show a strong dependency of $q_s$ on the collision energy, 
$\sqrt s$,~\cite{shmc1,shmc2,CMSES}. In a shower based  event generator, there is a limit to the transverse momentum of parton emitted through QCD radiation of the order of 1~GeV.  By introducing such a cut, a certain part of soft emissions is excluded.  
\begin{figure} [hp]
\includegraphics[width=.49\linewidth]{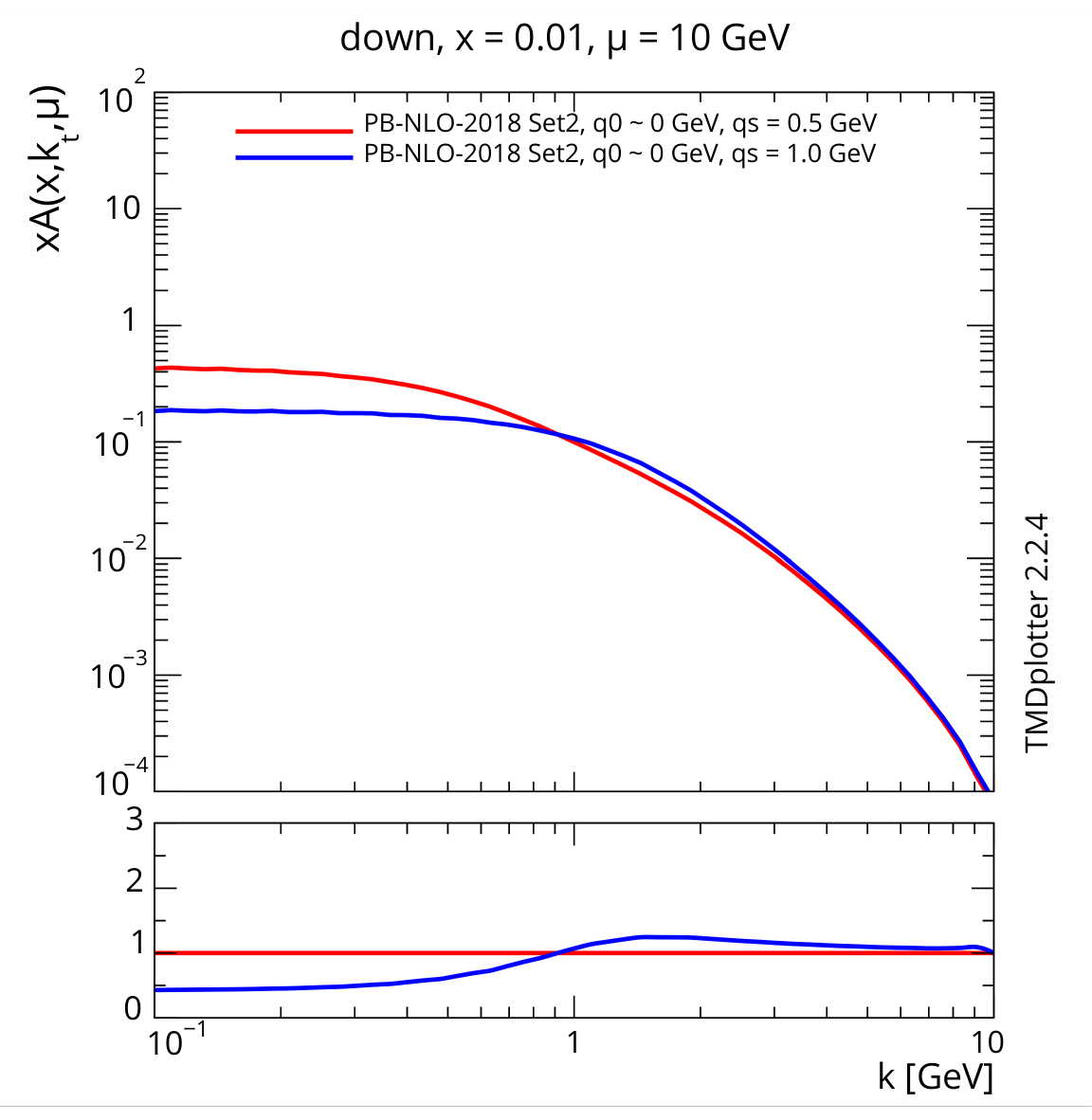}
\includegraphics[width=.49\linewidth]{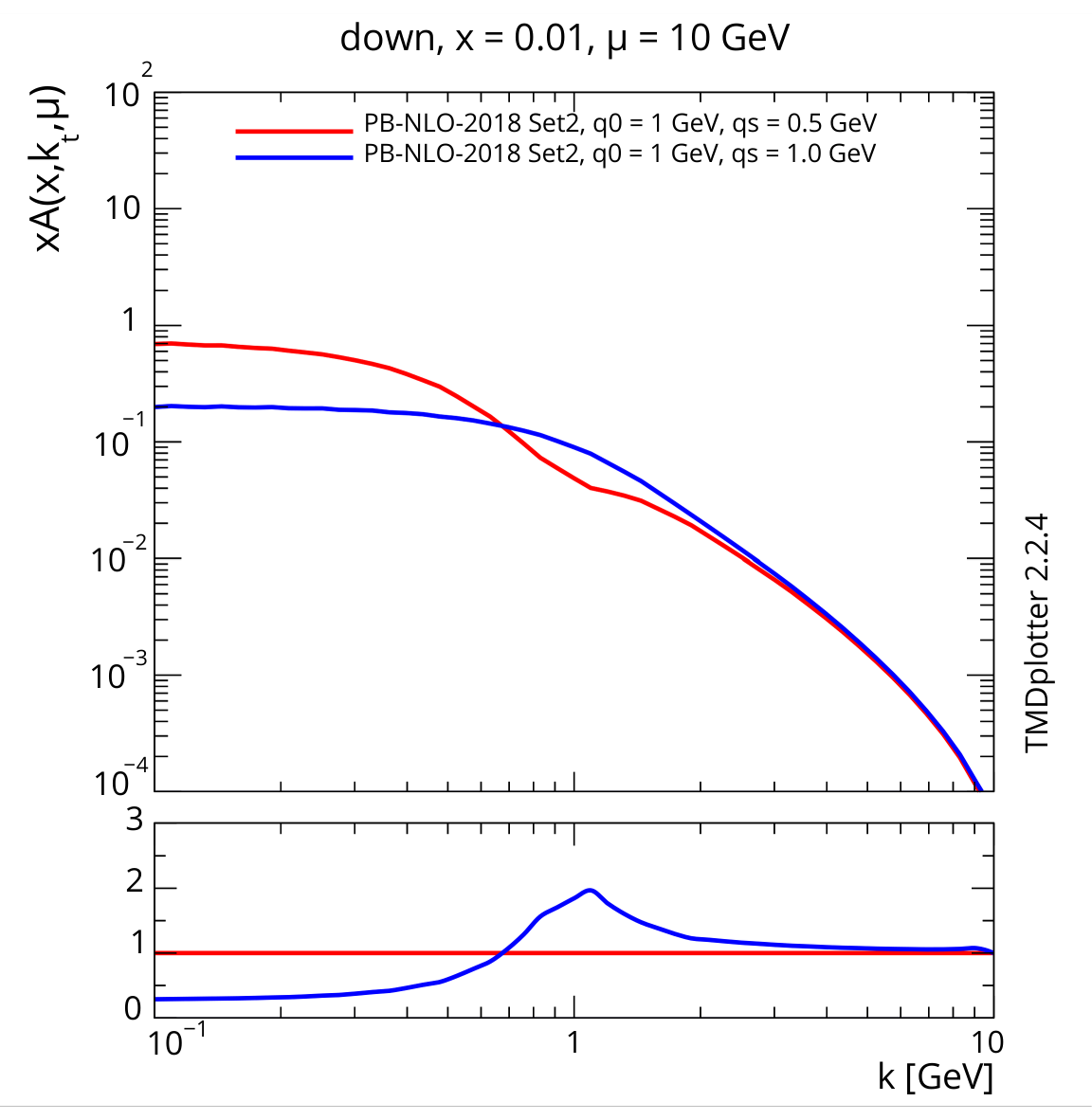}\\
\includegraphics[width=.49\linewidth]{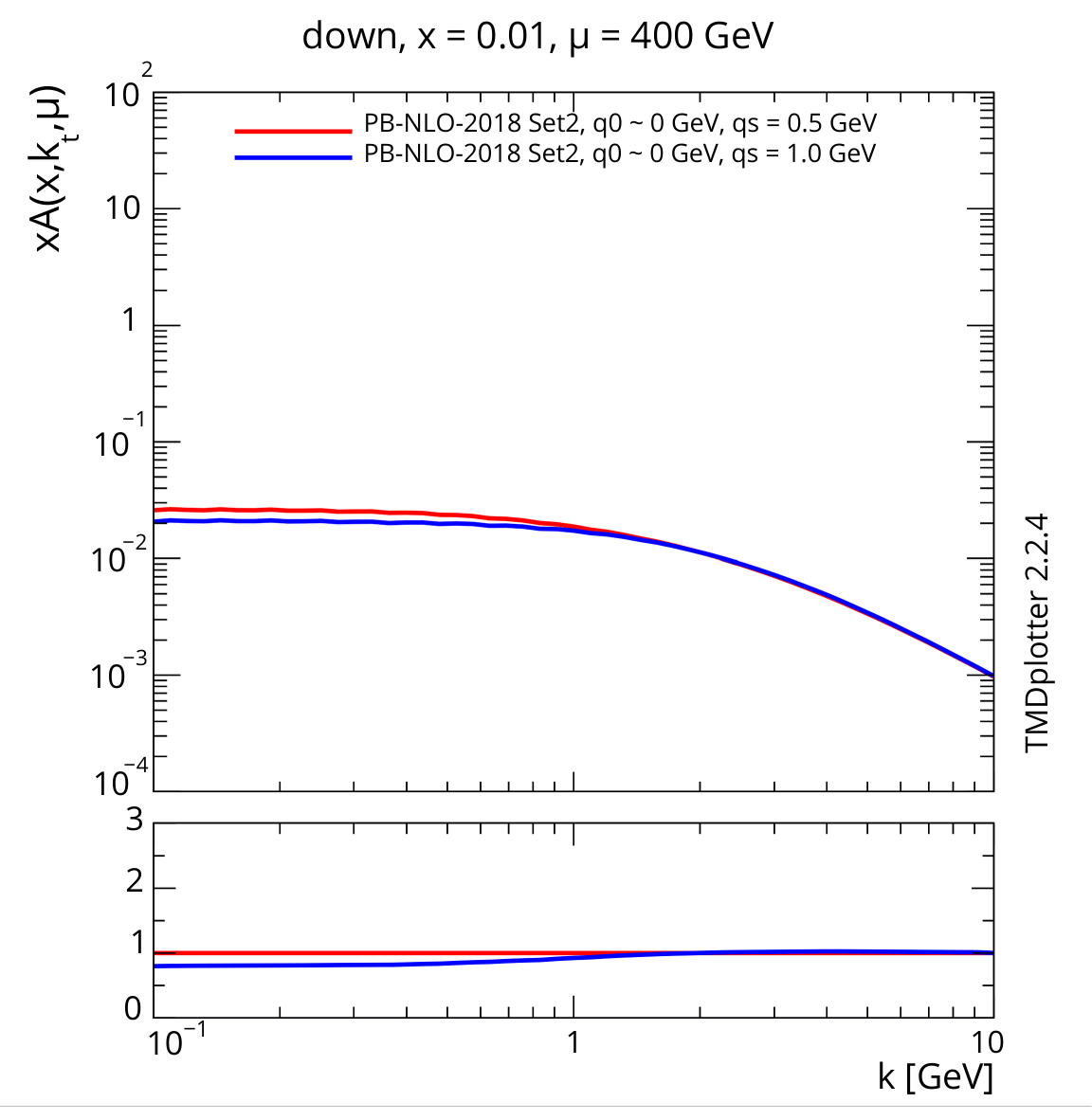}
\includegraphics[width=.49\linewidth]{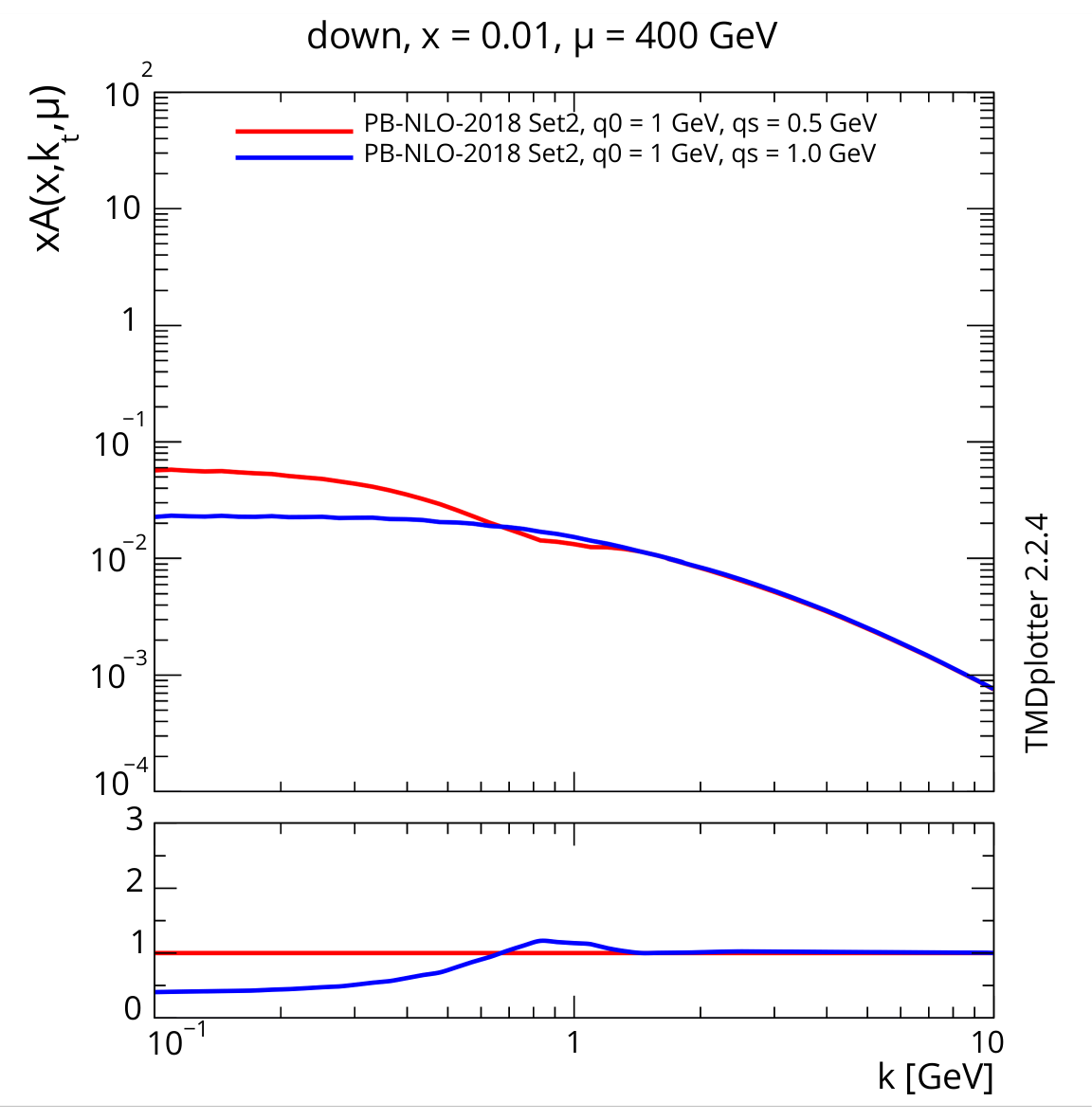}
\caption{Transverse momentum distributions for down quarks based on the PB-NLO-2018 Set2 TMD for two scale values: $\mu = 10$~GeV (up) and $\mu = 400$~GeV (down). The distributions are obtained for $x=0.01$ and presented for two values of the intrinsic-$k_{\rm T}$ width, $q_s = 0.5$~GeV (red) and  $q_s = 1.0$~GeV (blue) for two values of $q_0$: $q_0 \simeq 0$~GeV (left) and $q_0 = 1.0$~GeV (right).  The ratio plots show the ratios to the one for $q_s = 0.5$~GeV.}
\label{tmds}
\end{figure}

To cross-check our result and try to understand the energy scaling of $q_s$, we introduced a lower bound on the transverse momentum of the emitted parton in the branching, $q_0$, such that $q_{\rm \perp} > q_0$. Figure~\ref{tmds} shows a comparison of the transverse momentum distributions ($k_{\rm T}$) of down quarks based on PB-NLO-2018 Set2 for two values of the intrinsic-$k_{\rm T}$ width,  $q_s = 0.5$~GeV and $q_s = 1$~GeV, for two values of $q_0$: $q_0 \simeq 0$~GeV and $q_0 \simeq 1$~GeV. These sets are compared for two evolution scale values: $\mu = 10$~GeV and $\mu = 400$~GeV.  The plots are obtained using the graphical interface - TMDplotter\cite{tmdplotter1,tmdplotter2}. As expected, the intrinsic motion of the partons affects the lowest transverse momenta of the radiated partons. It can be seen that the difference between the $k_{\rm T}$-distributions with different $q_s$ increases with $q_0$ and decreases at larger scales, $\mu$, which means that there is a better sensitivity of the intrinsic-$k_{\rm T}$ width at larger $q_0$
and at smaller DY pair invariant masses (created from partons at small $\mu$). Thus, to check the energy scaling behaviour and get the best possible sensitivity, we determined the values of $q_s$ at different energies using the lowest pair mass bins (in case that there are measurements in several mass bins). If there is finer binning in the Z-peak region due to high signal and small background, as is the case with CMS measurement~\cite{CMS}, the Z-peak region was used to determine $q_s$. 
%
%============================= Fig. 1 ================================
\begin{figure} [h]
\begin{center}
\includegraphics[width=.75\linewidth]{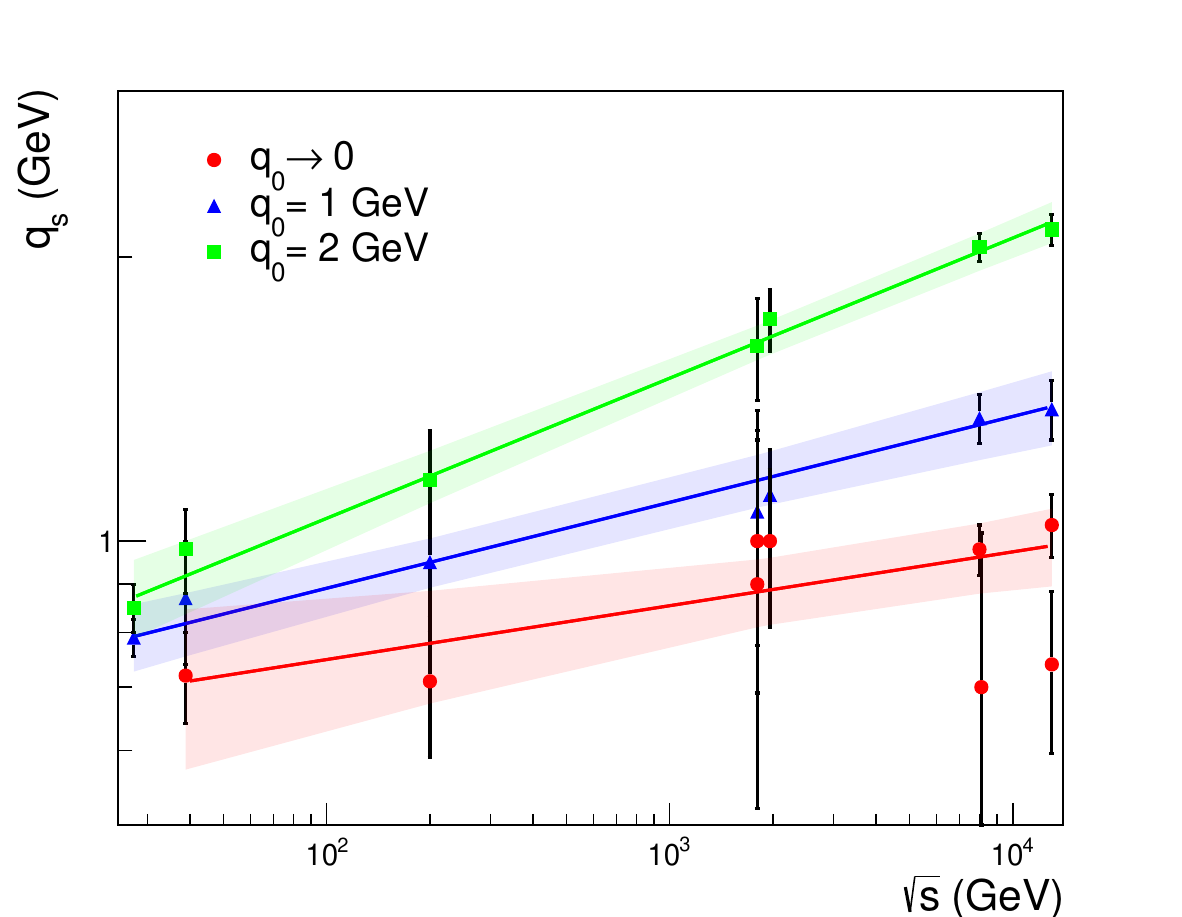}
\caption{Intrinsic-$k_{\rm T}$ width as a function of center-of-mass energy obtained by comparing experimental data and CASCADE3 predictions for three values of $q_0$: $q_0 \simeq 0$~GeV (red points), $q_0 = 1$~GeV (blue points) and $q_0 = 2$~GeV (green points). Lines and bands are the results of the fits and uncertainties explained in the text~\cite{ktpaper,pbescaling}. }
\label{enscaling}
\end{center}
\end{figure}
%============================= Fig. 1 ================================
%
Figure~\ref{enscaling}  shows the result of determining $q_s$ from data sets at different collision energies obtained in the procedure of minimisation of $\chi^2 (q_s)$ calculated by comparing the data with CASCADE3 predictions for three different values of cutoff parameters, $q_0$~\cite{ktpaper,pbescaling}. 
The $q_s (\sqrt s)$ dependence for each $q_0$ value is fitted with the function $q_s = a\cdot(\sqrt s)^b$ and the uncertainties of the fits at 95~$\%$ CL are obtained from the uncertainties of the parameters and presented as shaded regions. 
The figure shows that the intrinsic-$k_{\rm T}$ energy scaling is introduced by requiring a minimum transverse momentum at a branching as it is in shower based MC event generators. With this, the $z_{\rm M}$-parameter becomes constrained, $z_{\rm M} = 1- q_0/q'$, according to the angular ordering, and becomes equal to $z_{\rm {dyn}}$ by which the non-parturbative Sudakov form factor, $\Delta_a^{(NP)}$, is neglected. Since this contribution changes significantly with $q_0$, and since the slope of the $q_s(\sqrt s)$ fit increases with it, it can be concluded that the energy scaling behaviour of the intrinsic-$k_{\rm T}$ originates from the  omission of the non-perturbative soft contribution which interplays with intrinsic-$k_{\rm T}$.      

\label{sec:pythia}
\section{The intrinsic-$k_{\rm T}$ at 13~TeV in PYTHIA }

\begin{figure}[hp]
\includegraphics[width=.49\linewidth]{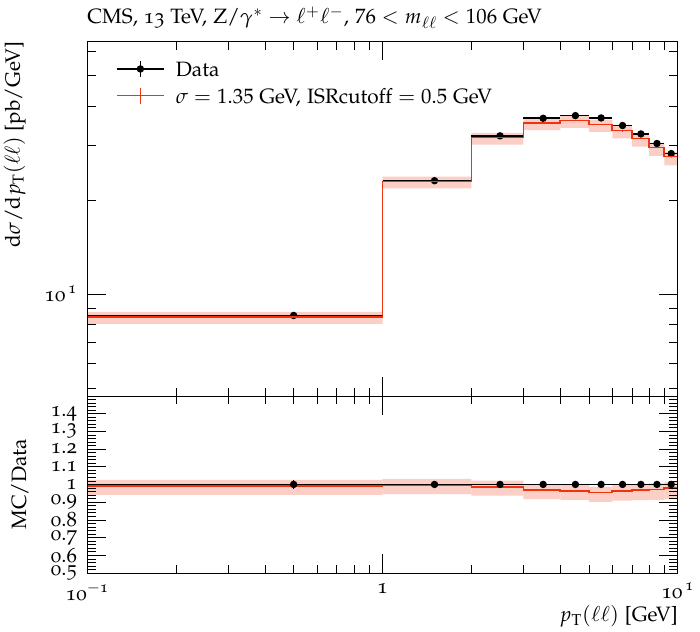}
\includegraphics[width=.49\linewidth]{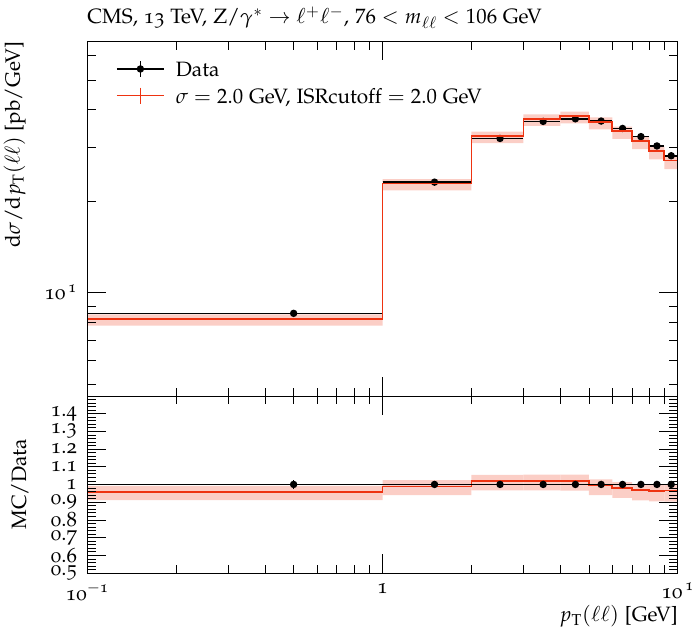}
\caption{Cross section as a function of DY pair transverse momenta in the Z-peak region measured from pp collisions at 13~TeV~\cite{CMS} compared to predictions from PYTHIA for two values of the ISR cutoff parameter. The value of the intrinsic-$k_{\rm T}$ width, $\sigma$,
obtained in the $\chi^2 (\sigma)$ minimisation procedure is indicated in each case.}
\label{sigmapythia}
\end{figure}
To cross check our observations and conclusions about the physical origin of the energy scaling behaviour of the intrinsic-$k_{\rm T}$ width obtained with CASCADE3, we analysed how distributions obtained from  PYTHIA8~\cite{pythia} event generator compare with experimental distributions from proton-proton
collisions at $\sqrt s= 13$~TeV~\cite{CMS} and extracted the intrinsic-$k_{\rm T}$ width the same way as we did in the case of CASCADE3 distributions. For this purpose we have used Monash tune~\cite{monash} modified so that the strong coupling in the shower is changed from $\alpha_s(M_{\rm Z}) = 0.1365$ as in the default setting to $\alpha_s(M_{\rm Z}) = 0.130$ to obtain as better as possible description of the data by the prediction.

In PYTHIA settings, a cutoff scale parameter which reduces soft contributions through the ISR  (similar to $q_0$ in PB) is denoted as $SpaceShower:pT0Ref$ and 
$BeamRemnants:primordialKThard$ 
 is the Gauss width, $\sigma$, of the intrinsic-$k_{\rm T}$ distribution (as $q_s/\sqrt{2}$ in the PB).
 
 Figure~\ref{sigmapythia} shows a comparison of data~\cite{CMS} and predictions obtained from PYTHIA with two values of the ISR cutoff parameter, 0.5 GeV  and 2.0 GeV, and with the intrinsic-$k_{\rm T}$ widths, $\sigma$ obtained as a minimum of $\chi^2(\sigma)$ for each of the cutoff parameters: $\sigma = 1.35$~GeV and $\sigma = 2.0$~GeV respectivelly. The red bands around PYTHIA predictions are obtained by varying the renormalization and factorization scales by factors of two up and down. The width $\sigma$ increases with the ISR cutoff parameter which confirms our findings obtained from the PB Method. 

\label{sec:pythia}
\section{Conclusion}

By properly treating the soft contributions, the PB Method enables the determination of the intrinsic-$k_{\rm T}$ width which does not depend on the invariant mass of the DY pair, nor on the center-of-mass collision energy, $\sqrt s$. The inclusion of soft gluons, in particular the non-perturbative Sudakov, is crucial for providing a $\sqrt s$-independent intrinsic-$k_{\rm T}$. The intrinsic-$k_{\rm T}$ contribution can be disentangled from the non-perturbative Sudakov one only by the appropriate treatment of non-perturbative processes achieved by the PB Method due to the sensitivity to non-perturbative TMD contributions. The PYTHIA results at 13 TeV confirm that the intrinsic-$k_{\rm T}$ width increases with the ISR cutoff scale.
 
 \vskip 0.5 cm 
\noindent 
{\bf Acknowledgments.} 
 I am grateful to the organisers of the international conference  Hadron Structure and Fundamental Interactions: from Low to High Energies 2024 for the opportunity to give a plenary talk on the presented results which are based on the collaborative work in the CASCADE group.  I wish to acknowledge all the colleagues for the very fruitful and extensive collaboration. 

 \vskip 0.5 cm 
\noindent 
{\bf Funding.} 
The results presented here are part of a national scientific project that has received
 funding from Montenegrin Ministry of Education, Science and Inovation.

\end{document}